\documentclass[12pt, twoside]{article}
\usepackage{a4wide,amssymb,cite}
\usepackage{epsfig}

\newcommand{\be}{\begin{equation}}
\newcommand{\ee}{\end{equation}}
\newcommand{\bea}{\begin{eqnarray}}
\newcommand{\eea}{\end{eqnarray}}
\newcommand{\vep}{\varepsilon}

\begin{document}

\begin{titlepage}

\rightline{December 2008}

\begin{centering}
\vspace{1cm}
{\Large  {\bf Flux compactifications and supersymmetry breaking in 6D gauged supergravity$^*$}}\\

\vspace{1.5cm}

 {\bf Hyun Min Lee}\\
\vspace{.2in}

{\it Department of Physics and Astronomy, McMaster University \\
Hamilton, Ontario L8S4M1, Canada.} \\
{\small (E-mail: hminlee@mcmaster.ca)}

\end{centering}
\vspace{2cm}

\begin{abstract}
\noindent
We review on a recent construction of the on-shell supersymmetric brane action for the codimension-two branes with nonzero tension in the flux compactification of a 6D chiral gauged supergravity.
On dimesionally reducing on 4D gauged supergravity for a new supersymmetric unwarped background with conical branes, 
we consider the modulus stabilization for determining the soft masses of the scalars 
localized on the branes and show that the bulk $U(1)_R$ provides a new mechanism for mediating the SUSY breaking.

\end{abstract}

\vskip 1cm

\vspace{5cm}
\begin{flushleft}

$^*~$Invited review for Modern Physics Letters A

\end{flushleft}
\end{titlepage}

\section{Introduction}	

Flux compactifications\cite{fluxcompact}, particularly in type IIB string theory context, have recently drawn plenty of attention due to the fact that most of moduli endowed from string theory can be stabilized due to internal fluxes in extra dimensions. On the other hand, a codimension-two brane in higher dimensional gravity theories has been a new arena to tackle the cosmological constant problem, because a nonzero tension of the codimension-two brane only generates a deficit angle of the internal space without gravitating along the 4D spacetime\cite{deficit}, which is called the {\it self-tuning} mechanism. 
In particular, in six-dimensional gravity theories where an embedded codimension-two brane is regarded as the visible 4D universe, the need of compactification of two extra dimensions naturally brings us into a flux compactification with codimension-two brane(s)\cite{navarro}. There is another possibility of having codimension-two branes in compact extra dimensions with a sigma model scalar field\cite{sigma}.

There has been a revived interest in the 6D chiral gauged supergravity\cite{NS} after the Salam-Sezgin unwarped solution 
with a sphere\cite{SS} was generalized to the warped solution with nonzero codimension-two brane tensions\cite{branesol}. 
In this model, in the presence of the $U(1)_R$ magnetic flux, the extra dimensions with axial symmetry are compactified on a manifold topologically equal to a wedged sphere at the pole of which nonzero brane tensions are located.
The merit of the 6D supergravity model consists in that the dilaton equation of motion picks up the 4D Minkowski space as a unique regular solution with 4D maximal symmetry due to a specific form of the bulk action guaranteed by the bulk supersymmetry(SUSY).
However, it turns out that the conical flat solution suffers from a quantized brane tension because of the flux quantization condition\cite{fluxquant} and the 4D curved solutions with a naked singularity have been shown to exist too\cite{curved}. Moreover, the 6D supergravity action and the brane tension without dilaton coupling respects the scale invariance, which in turn would lead to a problematic massless modulus in 4D effective theory. After the stabilization of the massless modulus, it is not guaranteed that the contribution of the modulus potential to the vacuum energy is made as small as the observed cosmological constant without fine-tuning. Nonetheless, the codimension-two brane world has shed much light on the new aspect of the cosmological constant problem that would be worth a further study.  

Here, we study a different aspect of the warped flux compactifications in the same 6D gauged supergravity. 
The general warped solutions with conical branes found in the 6D gauged supergravity turn out to break all the bulk SUSY
at the compactification scale for generic parameters of the solution. 
If SUSY is a solution to the gauge hierarchy problem, however, we need to keep 4D ${\cal N}=1$ SUSY unbroken much below the compactification scale at which the higher dimensional SUSY is broken to 4D ${\cal N}=1$ SUSY and the SM gauge couplings are unified. 
Among the general conical brane solutions, there is a particular brane solution, for which the warp factor is constant, i.e. two equal brane tensions are situated at the poles of a football.
In this case, it has been shown that the 4D ${\cal N}=1$ SUSY remains thanks to the brane-localized Fayet-Iliopoulos(FI) terms coming from the SUSY completion of the brane tension action\cite{leepa}. On the supersymmetric codimension-two brane, one can introduce the Minimal Supersymmetric Standard Model(MSSM) fields with a fixed coupling to the bulk fields as required by the bulk SUSY\cite{lee}. Since the 6D gauged supergravity has the $U(1)_R$ symmetry gauged, it allows us to introduce the $U(1)_R$ coupling of the brane multiplets on top of the bulk supergravity coupling. 
Assuming that the SUSY breaking hidden sector and the visible sector are localized at different branes for a sequestering of the SUSY breaking\cite{sequester}, the bulk multiplets, both gravity multiplet and $U(1)_R$ vector multiplet can mediate the hidden brane SUSY breaking to the visible brane. 
Since the $U(1)_R$ gauge boson mass should be smaller than the 4D Planck mass by the $U(1)_R$ gauge coupling, it is conceivable that when the visible brane fields have nonzero $R$-charges, the $U(1)_R$ mediation gives a dominant contribution to the soft masses. 

The paper is organized as follows. 
We first give a brief review on the construction of the {\it on-shell} SUSY brane action in 6D gauged supergravity.
Then, after the 4D reduction of the 6D flux compactification on a {\it supersymmetric} football, we obtain the 4D effective supergravity with gauged $U(1)_R$ and matter fields 
and discuss on the $U(1)_R$-mediated SUSY breaking with modulus stabilization taken into account in the 4D effective supergravity.

\section{Supersymmetric codimension-two branes}

The 6D chiral gauged supergravity\cite{NS} is composed of the 6D minimal gravity multiplet that are a gravity multiplet($e^A_M,\psi_M,B^+_{MN}$), and a tensor multiplet($\phi,\chi,B^-_{MN}$), as well as a vector multiplet($A_M,\lambda$), which is needed to gauge the $U(1)_R$ symmetry. The 6D gauged supergravity is in contrast to the 5D minimal gauged supergravity where the $U(1)_R$ symmetry is gauged by the graviphoton that corresponds to an auxiliary multiplet in 4D effective supergravity\cite{5dgauged}.
In order to cancel the 6D gravitational anomalies, one has to introduce the additional vector and/or hyper multiplets satisfying $244+n_V-n_H=0$ where $n_V,n_H$ are the numbers of vector and hyper multiplets, respectively. The simplest possibility of the anomaly cancellation with gauged $U(1)_R$ is to introduce only $n_H=245$ hyper multiplets containing neutral hyperinos but there also exist other anomaly-free models with non-abelian and/or abelian gauge groups\cite{anomalyfree1,anomalyfree2}. 
We assume that the bulk hyper multiplets do not affect the background geometry and they are assumed to be decoupled at high energy below the compactification scale\footnote{We note that the background geometry with nonzero hyperscalar VEV preserving 4D ${\cal N}=1$ SUSY was found in Ref.~\cite{hypers}. However, since it is a singular solution without conical branes, we don't consider this case in the paper.}.

The bosonic Lagrangian of the 6D gauged supergravity is as follows,
\bea 
{\cal L}_{\rm bulk}&=&e_6\Big(R-\frac{1}{4}(\partial_M\phi)^2-\frac{1}{12}
e^{\phi}G_{MNP}G^{MNP}
-\frac{1}{4}e^{\frac{1}{2}\phi}F_{MN}F^{MN}-8g^2 e^{-\frac{1}{2}\phi}\Big)
\eea
where the field strength tensors are $F_{MN}=\partial_M A_N-\partial_N A_M$ and 
$G_{MNP}=3\partial_{[M}B_{NP]}+\frac{3}{2}F_{[MN}A_{P]}$.
Here $g$ is the $U(1)_R$ gauge coupling and we have set the 6D fundamental scale to $M^4_*=2$.
It is remarkable that a positive dilaton potential occurs due to the gauging of the $U(1)_R$ symmetry.
This is comparable to the 4D gauged supergravity\cite{4dgauged} where a constant $U(1)_R$ FI term can lead to a nonzero potential for D-term inflation\cite{binetruy}. 

In the 6D gauged supergravity, by turning on the magnetic flux in the extra dimensions while setting the KR field to zero, 
Salam and Sezgin found a 4D Minkowski solution without warp factor 
and with extra dimensions compactified on a sphere\cite{SS}.
The solution has been recently generalized to the unwarped or warped 4D Minkowski solutions with nonzero
brane tensions situated at the conical singularities\cite{branesol}. The Lagrangian for a brane tension, ${\cal L}_{\rm brane}=-e_4 T\delta^2(y)$, however, has been considered to break the bulk SUSY explicitly, as $\delta {\cal L}_{\rm brane}=-e_4 \frac{1}{4}T({\bar\psi}_\mu \Gamma^\mu\vep+{\rm h.c.})\delta^2(y)$. 

In the gauged supergravity, the gravitino is charged under $U(1)_R$.
Thus, in order to make the brane tension action supersymmetric, we utilize the SUSY variation of the gravitino kinetic term.
First we note that varying the gravitino kinetic term gives rise to a piece of the gauge field strength as
$$
\delta{\cal L}_{\rm gravitino}=-\frac{i}{2}e_6\, g{\bar\psi}_M\Gamma^{MNP}\vep F_{NP}+\cdots.
$$  
In the above variation, we rewrite the gauge field strength in terms of the hatted one 
and a localized Fayet-Iliopoulos(FI) term parametrized by $\xi=\frac{T}{4g}$ as\cite{leepa}
\be
F_{mn}={\hat F}_{mn}+\xi\epsilon_{mn}\frac{\delta^2(y)}{e_2} \label{gaugefieldst}
\ee
where $\epsilon_{mn}$ is the 2D volume form.
Then, after {\it replacing} the gauge field strength with the hatted one in the bulk action and the fermionic SUSY transformations,
we can cancel the brane tension term by the variation of the gravitino kinetic term consistently.
We note that a $Z_2$ orbifold boundary condition with $\vep_R(y=0)=0$ must be imposed
to break half the bulk SUSY on the brane
and the strength tensors for the KR field appearing are also replaced with the modified ones\cite{leepa}.

Brane multiplets can be also accommodated by modifying further the field strength tensors and the SUSY transformations\cite{lee}.
For a chiral multiplet, a brane scalar with $R$ charge $r_Q$ gets a mass term, ${\cal L}_{\rm mass}=-4r_Qg^2|Q|^2e_4$, 
due to the $U(1)_R$ gauging while the fermionic partner of the brane scalar having an $R$ charge $r_Q-1$ is massless.
The extra component of the gauge field strength tensor picks up an additional correction\cite{lee},
\be
{\hat F}_{mn}=F_{mn}-(\xi+r_Q g|Q|^2)\epsilon_{mn}\frac{\delta^2(y)}{e_2}.
\ee
Moreover, the strength tensor for the KR field gets additional terms\cite{lee}:
\be
{\hat G}_{\mu mn}=G_{\mu mn}+(J_\mu-\xi A_\mu) \epsilon_{mn}\frac{\delta^2(y)}{e_2}
\ee
where $J_\mu$ is the Noether current of the brane matter multiplets,
and the one with 4D components only also gets modified too. Therefore, the Bianchi identities for the modified
field strength tensors are modified due to the brane fields\cite{lee}, which is in a similar spirit to the supersymmetric codimension-one brane action in heterotic M-theory\cite{horavawitten}.
The kinetic term for the brane chiral multiplet has a dilaton coupling as ${\cal L}_{\rm kin}=-e_4 e^{\frac{1}{2}\phi}(D^\mu Q)^\dagger D_\mu Q+\cdots$
while the kinetic term for a brane vector multiplet does not depend on the moduli.
Moreover, the brane $F$ and $D$ terms are ${\cal L}_F=-e_4 e^{\psi-\frac{1}{2}\phi}|F_Q|^2$ (with $F_Q=\frac{\partial W}{\partial Q}$ for a moduli-independent brane superpotential $W$, and $e^\psi$ the volume modulus) and ${\cal L}_D=-e_4\frac{1}{2}e^\phi D^2$, respectively. 
When there exist brane multiplets localized at the other brane(s), it is straightforward to generalize the modified field strength tensors by replacing the single delta term with more delta terms.

\section{Flux compactifications}

It has been shown that the general warped flat solutions\cite{branesol} are maintained in the presence of the FI terms\cite{leepa}. The general warped solution with 4D Minkowski space and internal axial symmetry  
takes the following form\cite{branesol},
\bea 
ds^2&=&e^{\frac{1}{2}\phi_0}\Big(W^2(r)
\eta_{\mu\nu}dx^\mu dx^\nu+R^2(r)(dr^2
+\lambda^2 \Theta^2(r)d\theta^2)\Big), \label{wmetric} \\
{\hat F}_{mn}&=&q e^{-\frac{1}{2}\phi} W^{-4} \epsilon_{mn},  \label{flux}\\
\phi&=&\phi_0+4\ln W ,
\eea with
\be
R={W \over f_0}, \ \  \ \Theta={r \over W^4},   
\ee
\be
W^4=\frac{f_1}{f_0}, \ \ f_0=1+\frac{r^2}{r^2_0}, \ \ \
f_1=1+\frac{r^2}{r^2_1},
\ee 
where $\lambda$,$\phi_0$ and $q$ are constant parameters, and the two radii $r_0$, $r_1$ are given by 
\be
r^2_0=\frac{1}{2g^2}, \ \ r^2_1=\frac{8}{q^2}. 
\ee

In the warped solution, the metric has two conical singularities,
one at $r=0$ and the other at $r=\infty$, which is at finite
proper distance from the former one. The singular terms coming from the deficit angles
at these singularities need to be compensated by brane tensions with the following matching conditions, 
\bea
T_1&=&4\pi (1-\lambda), \label{tension1}\\
T_2&=&4\pi\Big(1-\lambda\frac{r^2_1}{r^2_0}\Big).\label{tension2}
\eea 
After solving the gauge equation (\ref{gaugefieldst}) with two localized FI terms at the conical singularities, 
the quantization condition for the $U(1)_R$ gauge flux gets modified as
\be 
\frac{4\lambda
g}{q}=n-\frac{g}{2\pi}(\xi_1+\xi_2), \ \ n \in {\mathbf Z}
\label{quantcond} 
\ee 
where $\xi_i=\frac{T_i}{4g}(i=1,2)$. 
Even with eq.~(\ref{tension1})-(\ref{quantcond}), there is an undetermined dilaton constant $\phi_0$.

Simply by looking at the dilatino SUSY variation, the general warped solution breaks 
the bulk SUSY completely\cite{gravitino,lee}.
Thus, we focus on the football solution for which the dilaton is constant as $\phi=\phi_0$ and the gauge flux is given by ${\hat F}_{\rho\theta}=4g\epsilon_{\rho\theta}$ and the metric solution does not have a warp factor as
\be
ds^2=e^{\frac{1}{2}\phi_0}\Big(\eta_{\mu\nu}dx^\mu dx^\nu+\frac{r^2_0}{4}(d\rho^2+\lambda^2\sin^2\rho d\theta^2)\Big). 
\ee
In this case, we need to locate two codimension-two branes with equal tensions, 
$T_1=T_2=4\pi (1-\lambda)$, at the poles of the football. 
The gauge field strength is modified due to two equal FI terms localized at the poles as $A_\theta=-\frac{\lambda}{2g}(\cos\rho\mp 1)\pm \frac{\xi_1}{2\pi}$ with $\xi_1=\frac{T_1}{4g}$.
Then, the flux quantization condition (\ref{quantcond}) with $q=4g$ and $\xi_1=\xi_2=\frac{T_1}{4g}$ is satisfied for
the monopole number $n=1$ and arbitrary $\lambda$. Consequently, an arbitrary brane tension $T_1$ is allowed.
It has been shown that in the SUSY Killing equation, there occurs a cancellation between the spin connection and the gauge connection for the football solution as in the Salam-Sezgin solution so 
the football solution preserves 4D ${\cal N}=1$ SUSY\cite{leepa}. 
Therefore, the football solution corresponds to a {\it self-tuning supersymmetric} solution. 
After the brane-localized SUSY breaking, however, the self-tuning would not be guaranteed as usual in non-supersymmetric compactifications, because the effective localized FI terms are not proportional to the effective brane tension.

\section{The 4D effective supergravity with gauged $U(1)_R$}

Now we discuss on the low energy supergravity with brane multiplets by dimensionally reducing on 4D for the football geometry.
To that purpose, we take the ans\"atze for the 6D solution as
\bea
ds^2&=& e^{-\psi(x)}g_{\mu\nu}(x)dx^\mu dx^\nu + e^{\psi(x)} ds^2_2, \nonumber \\
\phi&=&f(x),  \\
{\hat F}_{MN}&=&\langle {\hat F}_{MN}\rangle+{\cal F}_{MN},   \nonumber  
\eea
where $\langle {\hat F}_{MN}\rangle$, ${\cal F}_{MN}$ are the VEV and fluctuation of the gauge field strength,
respectively, $ds^2_2$ is the 2D metric of the football solution and $f(x),\psi(x)$ are the scalar modes independent of the extra coordinates.
It has been shown that another constant scalar mode gets massive due to the flux and the non-constant scalar modes correspond to massive ones\cite{scalarpert}.
Then, by solving the 6D equations and the Bianchi identities for the modified field strengths\cite{lee}, 
we obtain 
\bea
{\hat G}_{\mu mn}&=&\Big(-b+4gA_\mu+\frac{J_\mu}{V}\Big)\epsilon_{mn}, \\
{\hat F}_{mn}&=&\Big(4g-\frac{r_Qg|Q|^2}{V}\Big)\epsilon_{mn},
\eea
where $b=-\frac{1}{2}{\cal B}_{mn}\epsilon^{mn}$ for the globally well-defined 
${\cal B}=B-\frac{1}{2}\langle A\rangle\wedge {\cal A}$ that satisfies 
$\delta_{\Lambda_0} (d{\cal B})$=0 for the background gauge transform $\Lambda_0$, 
and $V=\lambda \pi r^2_0$ is the volume of extra dimensions for the football solution.

The 4D effective supergravity is described by the K\"ahler potential $K$ and the superpotential $W$.
Including brane chiral multiplets $Q,Q'$ present at both branes 
and plugging the above solutions into the 6D action together with $e^f G_{\mu\nu\rho}=\epsilon_{\mu\nu\rho\tau}\partial^\tau\sigma$ 
and integrating over the extra dimensions,
we identify the K\"ahler potential as\cite{lee}
\bea
K&=&-\ln\Big(\frac{1}{2}(S+S^\dagger)\Big)-\frac{2\xi_R}{M^2_P} V_R \nonumber \\
&&-\ln\Big(\frac{1}{2}(T+T^\dagger-\delta_{GS} V_R)-Q^\dagger e^{-2r_Qg_R V_R} Q
-Q^{'\dagger} e^{-2r_{Q'} g_R V_R} Q'\Big)
\label{kahler}
\eea
where the Green-Schwarz parameter is $\delta_{GS}=8g_R$ and the coefficient of the $U(1)_R$ constant FI term is $\xi_R=2g_RM^2_P$  with $g_R=g/\sqrt{V}$ and 
the scalar components of the moduli superfields $S,T$ are given by
$$
S=e^{\psi+\frac{1}{2}f}+i\sigma, \quad T=e^{\psi-\frac{1}{2}f}+|Q|^2+|Q'|^2+ib
$$
Here $V_R$ is the $U(1)_R$ vector superfield.
The peculiar feature of our gauged $U(1)_R$ model coupled to the modulus $T$ is that there are both field-dependent and constant FI terms in the 4D effective supergravity. For the $U(1)_R$ gauge invariance of the K\"ahler potential, 
the $T$ modulus must transform as $\delta T=\frac{i}{2}\delta_{\rm GS}\Phi$  under the $U(1)_R$ gauge transformation, 
$\delta V_R=\frac{i}{2}(\Phi-\Phi^\dagger)$. 
This result is in contrast to the 6D ungauged supergravity where two extra dimensions are compactified on a torus orbifold\cite{faludlee}, e.g. $T^2/Z_2$.
Moreover, it has been shown\cite{lee} that the brane-localized superpotential is independent of the bulk moduli but the bulk-induced superpotential can depend on the moduli as will be discussed in the next section.
On the other hand, the gauge kinetic functions for the bulk and brane vector multiplets are
$f_R= S$ and $f_W = 1$, respectively\cite{lee}.

When the effective superpotential vanishes, from the determined K\"ahler potential (\ref{kahler}) with $f_R=S$, we obtain the 4D effective scalar potential only from the $U(1)_R$ D-term as
\be
V_0=\frac{1}{2}({\rm Re }S)D^2_R=\frac{2g^2_RM^4_P}{{\rm Re}(S)}\bigg[1-\frac{1-r_Q|Q|^2-r_{Q'}|Q'|^2}{{\rm Re}(T)-|Q|^2-|Q'|^2}\bigg]^2.
\ee
This scalar potential is consistent with the one derived directly from the 6D supersymemtric bulk-brane action\cite{lee}.
So, we have ${\rm Re}(T)=1$ and $|Q'|=|Q|=0$ stabilized at the SUSY minimum with a zero vacuum energy while ${\rm Re}(S)$ is undetermined. In the process of stabilizing the $T$ modulus, a large constant FI term, that is always present in 4D gauged supergravity, is cancelled by a field-dependent FI term coming from the bulk $U(1)_R$ flux.
The axionic part $b$ of the $T$ modulus is eaten up by the 4D component of the $U(1)_R$ gauge boson by a Green-Schwarz mechanism\cite{gs}. 
As expected for a massless chiral multiplet in 4D Minkowski space, the scalar partner of a massless fermion has a vanishing mass due to the cancellation between the brane mass term and the flux-induced mass term.

\section{Modulus stabilization and SUSY breaking}

In order to stabilize the $S$ modulus, we assume that the bulk gaugino condensates generate an $S$-dependent superpotential $W(S)$. This is possible because the gauge kinetic function of a bulk non-abelian gauge group is proportional to the $S$ modulus\cite{anomalyfree1} as for the $U(1)_R$ gauge kinetic function.
For instance, in the presence of the double gaugino condensates with matter fields decoupled and the Polonyi-type SUSY breaking on the hidden brane, the effective superpotential relevant for SUSY breaking and modulus stabilization is given by
\be
W= fQ'+\Lambda_1 e^{-\beta_1 S}+\Lambda_2 e^{-\beta_2 S}.
\ee 
We denote the resulting additional potential due to the gaugino condensation by $V_1=e^K(|D_SW|^2 K^{-1}_{SS^\dagger}-2|W|^2)/M^2_P$ where the $T$-modulus F-term has been included. 
Then, for $|\beta_1-\beta_2|\ll \beta_1$, the $V_1$ potential is minimized at a large ${\rm Re} (S)$.
When the $S$ modulus is stabilized only by $V_1$, $D_S W=0$ at the minimum so the vacuum energy would become negative.
Therefore, we need to lift the vacuum energy up to as small a positive value as observed by means of the $F$ and/or $D$ terms on the hidden brane.
The Polonyi-type superpotential on the brane leads to a nonzero $F$ term potential, $V_2=\frac{1}{{\rm Re}(S)}|F_{Q'}|^2$
with $F_{Q'}=\frac{\partial W}{\partial Q'}$. Here we note that $V_1+V_2$ comprises the total F-term potential of the model.
On the other hand, when there is a nonzero D-term on the hidden brane, the corresponding D-term potential is $V_3=\frac{D^2}{2({\rm Re}(T)-|Q|^2-|Q'|^2)^2}$.

Including the non-perturbative correction and the uplifting potentials, 
the total 4D scalar potential becomes
\be
V_{\rm tot}=V_0+V_1+V_2+V_3.
\ee
Then, $|Q|=0$ is still the minimum for $r_QD_R>0$. Assuming that the SUSY breaking field $Q'$ does not get a VEV, 
the minimum of ${\rm Re}(T)$ is shifted to
\be
{\rm Re}(T)=\frac{1+\frac{1}{2}\alpha D^2}{1-\frac{1}{2}\alpha {\rm Re} (T) V_1}; \quad \alpha\equiv\frac{{\rm Re}(S)}{2g^2_R M^4_P}
\ee
where the right-hand side is a $T$-independent quantity.
The $S$ modulus is also determined approximately by $D_S W=0$ and it is shifted a bit 
due to the $S$-modulus dependence of the hidden brane F-term.

After fixing all the moduli at the zero vacuum energy, we find that there is no gravity or modulus mediation at all but the $U(1)_R$ D-term is the only source of the tree-level SUSY breaking for a brane scalar. The soft mass of a scalar field in the visible brane is given by
\bea
m^2_Q=r_Q\,g_R D_R|_{Q=0} \label{gscalarmass}
\eea
where
\be
D_R=-\frac{1}{2g_RM^2_P}\Big(V_1+2V_2+V_3\Big).
\ee
From the result (\ref{gscalarmass}), a brane scalar with $r_Q=0$ has a vanishing mass at tree level.
Due to the functional form of the scalar potential, which is $V=V({\rm Re}(T)-|Q|^2)$ for $r_Q=0$, it is understood that the minimization condition, $\partial_T V=0$, is equivalent to a vanishing soft mass, $V_{Q^\dagger Q}=0$.

Up to now, we have ignored the effect of nonzero scalar VEVs of chiral superfields participating in the gaugino condensates.
In order to get the gaugino condensates, we would need nonzero VEVs of $R$-charged scalars because the $S$ modulus does not transform under the $U(1)_R$. 
Thus, the scalar VEVs would also give a non-vanishing contribution to the F-term potential and the $U(1)_R$ D-term. However, as far as the scalar VEVs are stabilized at values smaller than the 4D Planck scale dominantly by the global SUSY conditions for F-terms, the scalar VEVs would affect our result little\cite{choilee}.

When the hidden D-term vanishes, i.e. $V_3=0$, for an ignorable $U(1)_R$ D-term and $V_1\simeq -2m^2_{3/2}$ with $m_{3/2}=|e^{K/2}W|$, the zero vacuum energy condition becomes $V_2\simeq -V_1\simeq 2m^2_{3/2}$. Therefore, the soft mass of an $R$-charged visible scalar becomes
\be
m^2_Q\simeq -r_Q m^2_{3/2}. 
\ee  
Then, for a positive scalar mass squared, the $R$ charge of the brane scalar must be negative.
Moreover, in order to avoid unacceptable flavor violations, the $R$-charges should be family-independent.
Thus, when the $R$-charges are negative and family-independent, we can regard the $U(1)_R$ mediation as a dominant source for the SUSY breaking to solve the problem of negative slepton soft mass squareds in anomaly mediation. 

If the $U(1)_R$ symmetry were anomaly-free, the $\sum_Q (r_Q-1)=0$ condition would require that some of the scalars have
positive $R$-charges, i.e. negative soft mass squareds. Therefore, we would need to extend the MSSM fields with SM singlets\footnote{See, Ref.~\cite{u1ranom}, for the anomaly cancellation with SM non-singlets.} and/or consider an anomalous $U(1)_R$.
Since the bulk Green-Schwarz term required to cancel the 6D reducible anomalies also leads to 4D $U(1)_R$ anomalies on the flux compactification\cite{lee}, massless modes of bulk fermions give nonzero anomaly contributions for the $U(1)_R$. 
Thus, it is conceivable that the $U(1)_R$ symmetry appears anomalous in the low-energy theory and the additional anomalies of the MSSM fermions localized on the brane can be cancelled by the brane-localized Green-Schwarz terms with the $T$-modulus dependence\cite{choilee}. The detailed SUSY spectrum depends on the anomaly cancellation conditions and it is possible to find a model where the SM mixed anomalies are cancelled by a Green-Schwarz mechanism for family-independent negative $R$-charges for all squarks and sleptons\cite{choilee}.
The detailed SUSY spectrum for the MSSM fields and the $U(1)_R$ phenomenology will be presented in a forthcoming publication\cite{choilee}.

\section{Conclusion}

We have given a review on the flux compactification with SUSY codimension-two branes in 6D chiral gauged supergravity and presented the $U(1)_R$ mediation as a dominant source of SUSY breaking in 4D effective supergravity with gauged $U(1)_R$.
We have pointed out that the localized FI terms accompanying the brane tensions modify the gauge potentials at the branes such that the football solution preserve 4D ${\cal N}=1$ SUSY. Although the $T$ modulus gets heavy due to the gauge flux, in order to make the low-energy theory consistent with the $U(1)_R$ invariance, it is important to keep the $T$ modulus as well as the $U(1)_R$ gauge boson in the 4D effective theory. The $S$ modulus remains massless but it can be stabilized by bulk gaugino condensates in the 4D effective theory level. After the moduli are stabilized at the Minkowski vacuum in the presence of a hidden sector SUSY breaking,
we have shown that there is no gravity or modulus mediation but the $U(1)_R$ mediation provides a dominant source for the scalar soft masses in the visible sector.

\section*{Acknowledgments}
The author would like to thank Adam Falkowski for an early input on the SUSY brane action and Antonios Papazoglou for various collaborations on codimension-two brane models.
This work is supported by the research fund from the Natural Sciences and Engineering Research Council (NSERC) of Canada.

\end{document}